\documentclass[letter]{aa} 
\usepackage{graphicx}
\usepackage{txfonts}
\newcommand{\micron}{$\mu$m}

\newcommand{\editone}[1]{ #1}

\begin{document}

\title{Hydrogenated amorphous carbon grains as an alternative carrier of the 9$-$13{\micron} plateau feature in the fullerene planetary nebula Tc~1}

\titlerunning{HAC-like grains as a possible carrier of the 12\,{\micron} plateau emission}
\authorrunning{G{\'o}mez-Mu{\~n}oz et al.}

\author{M.~A. G{\'o}mez-Mu{\~n}oz\inst{1,2}
          \and
          D.~A. Garc{\'\i}a-Hern{\'a}ndez\inst{1,2}
          \and
          R. Barzaga\inst{1,2}
          \and
          A. Manchado\inst{1,2,3}
          \and
          T. Huertas-Rold{\'a}n\inst{1,2}
          }

\institute{
        Instituto de Astrof{\'\i}sica de Canarias, E-38205 La Laguna, Tenerife, Spain \\ \email{magm@iac.es, agarcia@iac.es}
        \and
        Departamento de Astrof{\'\i}sica, Universidad de La Laguna, E-38206 La Laguna, Tenerife, Spain
        \and
        Consejo Superior de Investigaciones Cient{\'\i}ficas (CSIC), 28006 Madrid, Spain
             }

\date{Received 22 December 2023; Accepted 05 February 2024}

\abstract{
  Fullerenes have been observed in several astronomical objects since the discovery of C$_{60}$ in the mid-infrared (mid-IR) spectrum of the planetary nebula (PN) Tc~1. It has been suggested that the carriers of the broad unidentified infrared (UIR) plateau features, such as the 9$-$13\,{\micron} emission feature (12\,{\micron} hereafter), may be related to the formation of fullerenes. In particular, their carriers have been suggested to be mixed aromatic or aliphatic hydrocarbons such as hydrogenated amorphous carbon (HAC-like hereafter) grains. For this study, we modeled the mid-IR emission of the  C$_{60}$-PN Tc~1 with a photoionization code, including for the first time the laboratory optical constants ($n$ and $k$ indices) of HAC-like dust at 300 K. Interestingly, we find that the broad 12\,{\micron} plateau feature in Tc~1 is well reproduced by using a distribution of canonical HAC grains, while at the same time they provide an important fraction of the IR dust continuum emission and are consistent with the other UIR features observed (e.g., the broad 6$-$9\,{\micron} plateau feature). This finding suggests that HAC-like grains may be possible carriers of the 12\,{\micron} plateau feature, being likely related to the fullerene formation mechanism in PNe.  More laboratory experiments, to obtain the optical constants of HAC-like dust with several structures or a composition at different physical conditions, are strongly encouraged -- that is, in order to extend this pilot study to more fullerene PNe, and to unveil the details of fullerene formation and of the potential carriers of the elusive UIR plateau features.
  }


\maketitle
%


\section{Introduction}

Planetary nebulae (PNe) represent the late stages in the evolution of low- and intermediate-mass stars (LIM; $\sim$1$-$8\,M$_{\sun}$), the majority of stars in the Universe. On their way to the PN phase, LIM stars experience a strong mass loss during the preceding asymptotic giant branch \citep[AGB; see e.g.,][for a review]{herwig2005} phase and they chemically enrich the surrounding interstellar medium (ISM). The AGB stars are also major suppliers of dust grains and molecular species that are routinely seen in the local Universe \citep[e.g.,][]{Ferrarotti2006}, which makes them (and their subsequent evolutionary stages) very important for our fundamental understanding of the enrichment and chemical composition of the ISM.

Fullerenes, among the most resistant and stable three-dimensional molecules that are only formed by C atoms, such as C$_{60}$ \cite[first discovered in the laboratory by][]{Kroto1985}, were supposed to be widely spread in space for decades.
The presence of fullerenes in astrophysical environments was under debate until the infrared (IR) signatures of the C$_{60}$ and C$_{70}$ fullerenes were unambiguously detected in the \textit{Spitzer} mid-IR spectrum of the young PN Tc~1 \citep{Cami2010}. After the first detection of fullerenes in a PN, the presence of C$_{60}$ was found in several other astrophysical objects such as reflection nebulae \citep{Sellgren2010}, additional PNe \citep{GarciaHernandez2010,GarciaHernandez2011b,GarciaHernandez2012, Otsuka2014},
 peculiar R Coronae Borealis (RCB) stars \citep[][]{GarciaHernandez2011a}, post-AGB stars \citep{Zhang2011,Gielen2011},
 and Herbig Ae/Be stars \citep{Arun2023}, among others, by its four
strongest mid-IR emission features at $\sim$7.0, 8.5, 17.4, and 18.9\,{\micron}. However, the formation route of fullerenes in such H-rich objects, similar to the majority of PNe, is not well understood.
Nowadays, the most suitable routes for the formation of fullerenes are\footnote{We note that, more recently, \citet{Bernal2019} proposed the shock heating and ion bombardment induced processing of SiC grains as an alternative route toward fullerenes in the ISM, but such a strong energetic process seems to be unlikely to happen in the circumstellar environments of fullerene PNe where the atomic nebular emission lines can be well explained by photoionization and shocks seem to be unimportant for fullerene formation \citep[see e.g.,][]{GarciaHernandez2012}.}: i) the photochemical processing of hydrogenated amorphous carbon (HAC) grains or similar mixed aromatic or aliphatic
hydrocarbons \citep[i.e., same chemical composition but a different internal structure; hereafter HAC-like;][]{GarciaHernandez2010} and ii) the photochemical processing of large polycyclic aromatic hydrocarbons \citep[PAHs;][]{Berne2012,Murga2022}.

The C$_{60}$ fullerenes are mainly detected toward PNe, and fullerene PNe are very young low-excitation ($T_\mathrm{eff}\sim$30\,000--45\,000\,K) C-rich objects, which evolved from low-mass ($\sim$1$-$3\,M$_{\sun}$) progenitors \citep{GarciaHernandez2012, Otsuka2014}. Remarkably, the IR Spitzer spectra ($\sim$5-38\,{\micron}) of the fullerene-rich PNe are dominated by aliphatic hydrocarbon-rich dust emission features (superimposed on the underlying and featureless dust continuum emission), showing several broad unidentified IR (UIR) plateau  emission features at 6$-$9 (hereafter 7\,{\micron}), 9$-$13 (hereafter 12\,{\micron}), 15$-$20, and 25$-$35\,{\micron}  \citep[e.g.,][]{GarciaHernandez2012}. Characteristic aliphatic discrete features are located at 3.4 and 6.9 $\mu$m, arising from symmetric and asymmetric C-H stretching and bending modes of methyl and methylene groups attached to aromatic rings, respectively \citep[see e.g.,][]{Kwok2011}. The identification of the carriers of the broad (and discrete) UIR emission features widely observed in space (toward Solar System bodies, circumstellar envelopes of PNe, diffuse ISM, and remote galaxies, among others) is a long-standing problem in astrophysics \citep[see e.g.,][for a review]{kwok2016}. Thus, the potential identification of the carrier(s) of any of the broad UIR plateau emission features in fullerene PNe would impact very different astronomical fields.

\editone{Nano- and micro-sized particles in the ISM and circumstellar environment play an important role in astrophysical processes that lead to the formation of such exotic UIR emissions seen in the spectra of C-rich PNe and other astrophysical objects. The optical constants, which are the complex refractive indices or the complex dielectric function, of different materials (e.g., graphite, silicates, SiC, MgS, and HAC) are commonly used, in conjunction with radiative transfer codes, to explain the origin of the IR emission in  astronomical spectra \citep[e.g., HACs, soots, SiC, and other hydrocarbon materials; see ][respectively]{BernardSalas2012,Gavilan2016,GomezLlanos2018,Dubosq2023}. Fortunately, the latter can be done thanks to the databases\footnote{ For example, the Jena database (\url{https://www2.mpia-hd.mpg.de/HJPDOC/index.php}) and the refractive index database (\url{https://refractiveindex.info/}).} dedicated to compile optical properties of analog materials of
cosmic dust in the wavelength range from the UV to the far-IR. }

In particular, the broad UIR plateau emission feature at 12\,{\micron} was first observed in the low resolution spectrometer of the IR astronomical satellite (LRS IRAS) spectra by \citet{Cohen1985}, and its carrier's identification has been a long outstanding problem.  \citet{Cohen1985} attributed  the carrier to PAHs, while \citet{Blanco1988} and \citet{Buss1990} attributed both  the 7 and 12\,{\micron} plateau features to a mixture of large PAHs and HAC grains.  Later, \citet{Kwok2001} specifically suggested that the 7 and 12\,{\micron} plateaus are superpositions of in-plane and out-of-plane bending modes of aliphatic side groups. Despite these earlier suggestions, the broad UIR emission at  12\,{\micron} has been generally attributed, in the literature,  to SiC \citep[$\alpha$-SiC; e.g.,][]{Speck2009,Jones2023}. \citet{GarciaHernandez2012b} argued against the SiC identification of this feature in fullerene PNe because the spectral characteristics (position and shape) \citep[see Table 4 in][]{GarciaHernandez2012} of the  9$-$13\,{\micron} emission in C$_{60}$-PNe significantly differ from those of the SiC 11.5\,$\mu$m feature seen in the preceding AGB phase. This may suggest that the carrier could be different in AGBs and in fullerene PNe or even that the identification of SiC in AGB stars could be revisited.

The PN Tc~1 is the prototypical fullerene-rich PN \citep[e.g.,][]{Cami2010,Aleman2019}, being one of the best objects to study because of its simple round-elliptical morphology, the available high-quality and high-resolution IR \textit{Spitzer} spectra, and the lack of PAH emission at 6.2, 7.7, 8.6, and 11.3 $\mu$m \citep[][]{Cami2010}, which could strongly contribute to the fullerene and broad (and discrete) UIR emission features. In this Letter, we present an analysis of the high-quality \textit{Spitzer} spectrum of PN Tc~1 in comparison with a photoionization model that includes, for the first time, the emission of HAC-like dust grains as measured in laboratory, showing that HAC-like grains are a possible explanation, different from the often assumed SiC, for the broad 12\,{\micron} plateau emission feature. 


\section{Infrared spectrum of the C$_{60}$-PN Tc~1} \label{sec:obs_data}

\begin{figure*}
    \centering
    \includegraphics[width=0.65\textwidth]{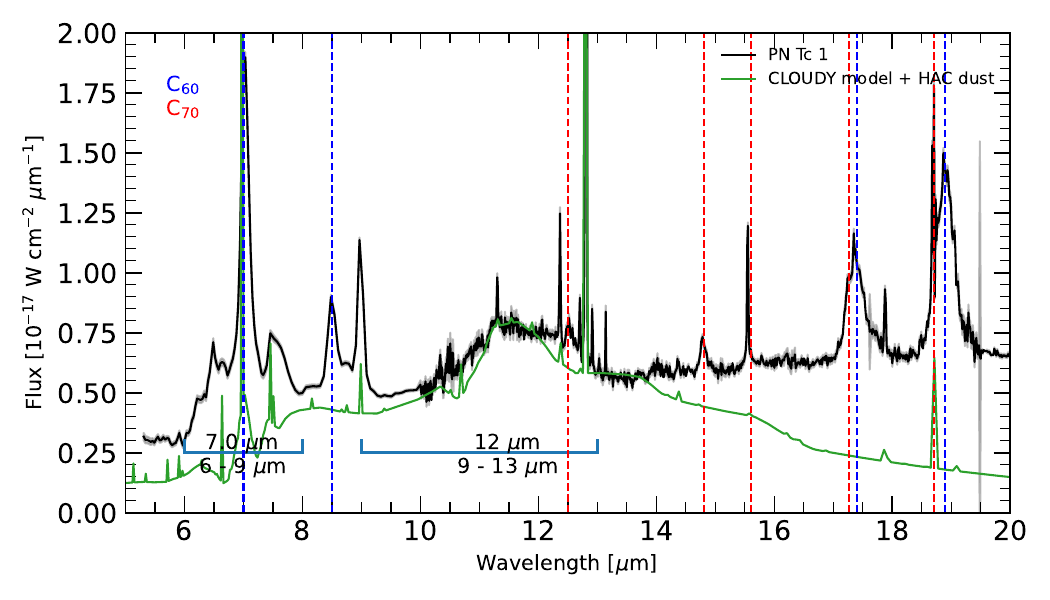}
    \caption{Spitzer mid-IR spectrum of the PN Tc~1 (black line) compared with the best photoionization model spectrum including HAC-like dust grains (green line). The C$_{60}$ (vertical dashed blue line) and C$_{70}$ (vertical dashed red line) emission bands and UIR plateau emission features are indicated. We note that the narrow emission features seen in the best model spectrum (green) are just atomic nebular emission lines, whose detailed modeling is out of the scope of the present work.}
    \label{fig:spitzer_model_tc1}
\end{figure*}

We downloaded all the available mid-IR spectra from the \textit{Spitzer} IR Spectrograph (IRS) \citep{Werner2004,Houck2004} Heritage archive of the PN Tc~1 for which high-quality -- a signal-to-noise ratio  (S/N) $>$ 50 -- short-low (SL; 5.2$-$14.5\,{\micron}), short-high (SH; 9.9$-$19.6\,{\micron}), and long-high (LH; 18.7$-$37.2\,{\micron}) spectra were available (Program Identifier 3633) \citep[see below and][for more data reduction details]{PereaCalderon2009}.
We used the package {\sc smart}\footnote{SMART was developed by the IRS Team at Cornell University and is available through the Spitzer Science Center at Caltech.} \citep[v8.2.9;][]{Higdon2004} to process the extracted wavelength and flux calibrated 1D spectra for each nod position, that is, to clean for bad data points, spurious jumps, and glitches and to combine and merge into one single spectrum per module (SL, SH, and LH). In order to get the final merged spectrum, we scaled the flux density of the SH spectra to match the overlapping wavelength region in the LH. Then, we scaled the SL spectra to match the overlapping wavelength region to the SH spectra. Finally, we scaled the flux density of the resulted combined spectra to that of the Wide-field IR Survey Explorer (WISE) band 4 ($\lambda_\mathrm{c}$=22\,{\micron}) photometric value \citep[similarly to what was done in][]{Otsuka2014}. The resulting final Tc 1 \textit{Spitzer} IR spectrum is shown in Fig.~\ref{fig:spitzer_model_tc1}, where the emission features such as the C$_{60}$ and C$_{70}$ IR bands, and the broad 7 and 12\,{\micron} plateau emission features are indicated.

\section{Cloudy simulations} \label{sec:analysis}

To the best of our knowledge, the only previous attempt to model the 7 and 12\,{\micron} plateau emission features seen in the \textit{Spitzer} spectrum of the C$_{60}$-PN Tc~1 (Fig.~\ref{fig:spitzer_model_tc1}) was made by \citet{BernardSalas2012}. They used the \citet{Jones2012} theoretical absorption coefficients (\textit{Q$_{abs}$}) of 3\,nm HAC particles with a different H content (H/C=0.23, 0.29, and 0.35) in conjunction with a 200\,K blackbody \citep[see Fig.7 in][]{BernardSalas2012}. They could reproduce the 7\,{\micron} plateau emission feature -- as seen in the Tc 1 continuum-subtracted \textit{Spitzer} spectrum -- with some degree of success, depending on the atom hydrogen fractions of the HAC particles. However, they could not reproduce the Tc 1 12\,{\micron} plateau emission feature. In particular, a too strong 11.3\,{\micron} feature was theoretically predicted for all H/C ratios, and they did not study or analyze the HAC contribution to the IR dust continuum emission. \citet{BernardSalas2012} concluded that their approximate reproduction of the 7\,{\micron} plateau emission feature may imply that fullerenes are formed by photochemical processing of HACs. Below we explain how we used, for the first time, a detailed photoionization model of Tc 1 together with the laboratory optical constants of HAC-like dust in order to model the 7 and 12\,{\micron} plateau features and the underlying dust continuum emission.

\subsection{Modeling the 12\,{\micron} plateau emission}

\begin{figure}
    \centering
    \includegraphics[width=1.0\columnwidth]{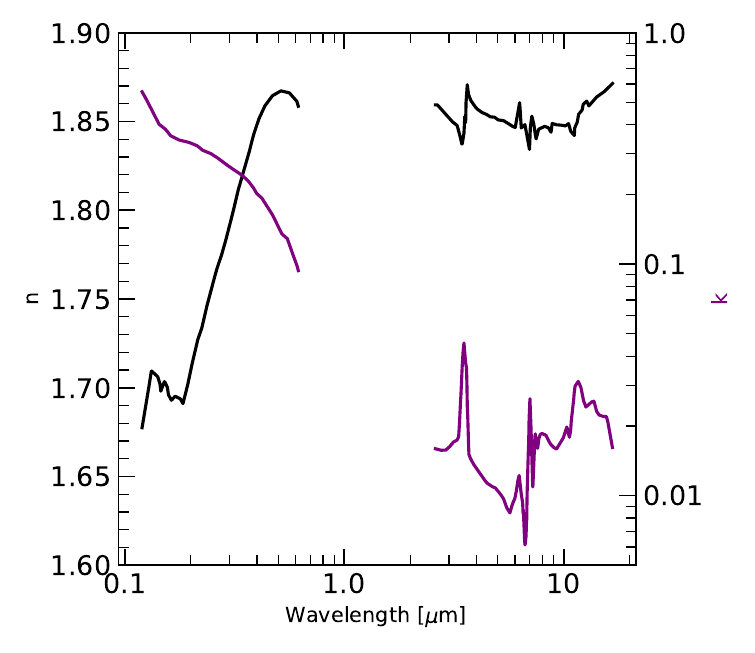}
    \caption{Laboratory average values of the complex refractive indices $n$ and $k$ (black and purple lines, respectively) of HAC-like particles at 300\,K and with a hydrogen concentration of H/C$\simeq$0.35 (Prof. W. Duley, private communication; see text for more details). \editone{The errors of the complex refractive indices were measured to be $\sim$20\% \citep[see e.g.,][]{Duley1984}}.}
    \label{fig:refractive_index}
\end{figure}

We used the stellar parameters and nebular abundances predicted in the detailed photoionization model developed by \citet{Aleman2019}\footnote{\citet{Aleman2019} developed a detailed photoionization model to obtain accurate stellar and physical parameters of the star and the nebula, respectively, and in which they assumed graphite in order to model the IR dust continuum emission. It is available at \url{https://github.com/Morisset/Tc1}} (without graphite as the IR dust continuum source; see below) as input in {\sc cloudy} \citep[v22.02;][]{Ferland2017} to model the mid-IR emission of the C$_{60}$-PN Tc~1 by using, for the first time, the laboratory optical constants of HAC-like dust.
We note that a change in the dust composition does not significantly change the emission line and H$\mathrm{\beta}$ fluxes predicted in the Aleman et al. model (a maximum flux difference of $\sim$5\% is obtained), and so the stellar and nebular parameters and chemical abundances are unaffected.
The laboratory optical constants, that is, the complex refractive indices $n$ and $k$ of HAC-like particles were obtained from Prof. W. Duley \citep[private communication; see][for a detailed description of the laboratory measurements]{Duley1984}, covering a spectral range from 0.1 to 17\,{\micron} with a gap between 0.65 to 2.5\,{\micron} \editone{due to a limitation of the laboratory instrumentation to measure over this wavelength range} (Fig.~\ref{fig:refractive_index}). The former are
\editone{the only available laboratory optical constants, prepared under simulated interstellar conditions, covering a sufficient range from the UV to the mid-IR}.
The laboratory $n$ and $k$ values were measured in HAC \editone{thin} films, \editone{$\sim$50$-$200\,nm} thickness, deposited at 300 K and with slightly different hydrogen concentrations \citep[H/C$\sim$0.3--0.4;][]{Duley1998,Duley2012b}. However, it should be noted that it is mostly the $k$ index that is sensitive to the H content; this is because the concentration of CH and CH$_{n}$ in the HAC samples determines the IR $k$. For simplicity, we adopted the mean $n$ and $k$ values (presumably with H/C$\sim$0.35), which are shown in Fig.~\ref{fig:refractive_index}.
Because {\sc cloudy} needs a minimum wavelength mesh ($\sim$0.001--96 $\mu$m), an extension of the $n$ and $k$ values to lower and upper wavelengths was carried out in a similar way as in \citet{GomezLlanos2018}; that is, we extrapolated the $n$ value to lower wavelengths on the basis of the BE1 amorphous carbon included in {\sc cloudy} \citep{Rouleau1991}, and for $k$ we extrapolated using a power law of slope 2, whereas for higher wavelengths we extrapolated $n$ as a constant value and $k$ using a power law of slope $-$1. For the gap between 0.65 to 2.5\,{\micron}, a linear extrapolation was used.

For the model, we compiled the HAC optical constants with the \texttt{grains} command inside {\sc cloudy}, using a canonical ISM grain distribution ($a_\mathrm{min}$=0.001\,{\micron} to $a_\mathrm{max}$=0.25\,{\micron}; 1$-$250\,nm\footnote{We note that equally good fits were obtained for different grain size distributions such as very small grains (VSG) or nanograins (1-15 nm), only big grains (15-250 nm), or the selected canonical ISM grain distribution (1-250\,nm).}, distributed in ten grain size logarithmic steps following a power of $-$3.5), and generated the opacities \citep[with the integrated Mie code for spherical grains; see][]{vanHoof2004} needed for the {\sc cloudy} simulations. We note that {\sc cloudy} takes into account the molar mass of the HAC (12.36\,g\,mol$^{-1}$), the band gap energy \editone{\citep[$E^\mathrm{ex}_\mathrm{gap}=1.27$\,eV;][]{Duley2012b}}, and the density of the material (1.5\,g\,cm$^{-3}$) \citep[see][]{Duley1998,Duley2012b}. A reduced $\chi^{2}_\mathrm{red}$ minimization was carried out for a grid of {\sc cloudy} models -- by varying the HAC grains' relative content -- to find a convergence between the model and the observed spectrum of Tc~1 between 10 and 14\,{\micron} (485 spectral points), yielding a $\chi^2_\mathrm{red}\simeq1.6$ assuming a 10\% error in the observations.
This resulted in a C abundance (in units of $\log$(C/H)) of $-$4.60 for the HAC-like grains, which is $\sim$3\% of the total C abundance (gas + dust phase abundance; C$\simeq-$3.08), a mean dust temperature across the nebula of $\sim$215\,K, and a dust-to-gas (D/G) ratio of 2.1$\times10^{-4}$.

Figure~\ref{fig:spitzer_model_tc1} also shows the best {\sc cloudy} model (green line) obtained from the $\chi^{2}_\mathrm{red}$ analysis applied to the observed spectrum of Tc~1. As it can be seen, the HAC-like grains at 300 K\footnote{We note that this value is close to the circumstellar dust grains' equilibrium temperature and the excitation temperature ($\sim$330\,K) of the C$_{60}$ emission in Tc 1 as derived by \citet{Cami2010}, who assumed that C$_{60}$ is attached to the dust grains.} provide a very good fit to the broad 12\,{\micron} plateau feature in Tc~1, while at the same time they fully reproduce the IR dust continuum emission from 9 to 14\,{\micron}. For comparison, $\alpha$-SiC dust grains were also used to similarly model the mid-IR spectrum of Tc~1, which resulted in a worst fit to the 12\,{\micron} feature (see Appendix~\ref{sec:ap_1}).
Interestingly, it turns out that HAC-like grains may be an alternative explanation of the broad 12\,{\micron} feature because it is very well reproduced with a simple spherical grain size distribution using the laboratory HAC optical constants at a deposition temperature close to the fullerene-dust temperature in Tc~1.

In addition, there is a small contribution to the observed 7\,{\micron} plateau feature, which is also clearly seen in the model spectrum. However, we still would need to include other emitters and/or an additional dust component to reproduce the 7\,{\micron} plateau feature (see below) in addition to the shape and rise of the underlying dust continuum emission at 5$-$9 and 14$-$17\,{\micron}, respectively.

\subsection{Modeling the 12\,{\micron} plateau emission and dust continuum}

Thus, in conjunction with the HAC-like grains as above, we added graphite dust grains separately as a possible additional dust continuum component (as assumed by \citet{Aleman2019}) and we again applied a $\chi^{2}_\mathrm{red}$ analysis to a grid of {\sc cloudy} models in order to find the best fit by varying the HAC grains and graphite relative contents in the 9 to 16\,{\micron} range, yielding a value of $\chi^2_\mathrm{red}\simeq$1.3. This resulted in C abundances of $-$3.74 and $-$4.90, which correspond  to $\sim$18.7\% and $\sim$1.3\% of the total C abundance (gas + dust phase) as well as D/G of 1.5$\times10^{-3}$ and 1.1$\times10^{-4}$ for graphite and HAC grains, respectively.

\begin{figure*}
    \centering
    \includegraphics[width=0.65\textwidth]{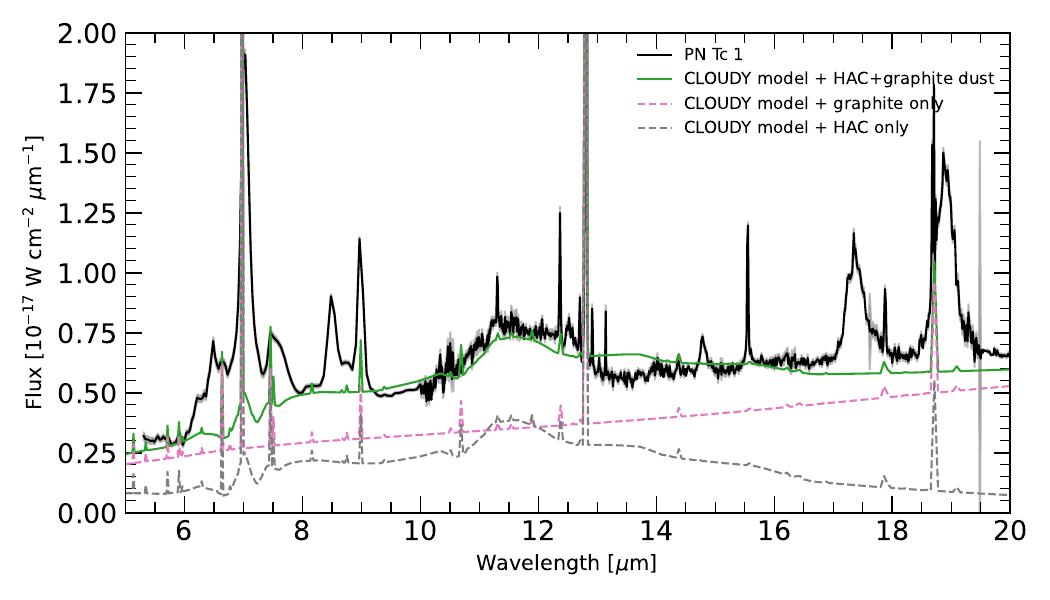}
    \caption{Spitzer mid-IR spectrum of the PN Tc~1 (black line) compared with the best photoionization model spectra for HAC-like (dashed gray line) and graphite (dashed magenta line) dust grains separately. The best photoionization model including HAC-like grains in conjunction with graphite dust grains (solid green line) is also shown for comparison.}
    \label{fig:tc1_dust_components_model}
\end{figure*}

We note that other carbon materials such as amorphous carbon (AC), generally used in the literature to model the IR dust continuum emission, provide worst fits because AC marginally contribute below 14\,{\micron}. Figure~\ref{fig:tc1_dust_components_model} displays the best photoionization model spectra for HAC-like grains and graphite, overplotted on the Tc 1 \textit{Spitzer} spectrum; the HAC-like and graphite IR emission contributions are also shown separately for comparison. This way, we can reproduce the Tc 1 IR spectrum well from $\sim$5 to 17\,{\micron} (both the 12\,{\micron} plateau feature and the dust continuum) with the HAC-like grains providing an important fraction ($\sim$20-40$\%$) of the dust continuum emission, as evaluated from the integrated IR fluxes due to only graphite and HACs with respect to the total integrated IR continuum flux at three different wavelengths (5.6, 9.6, and 15.2 $\mu$m) representative of the dust continuum emission. Again, for comparison, $\alpha$-SiC dust grains in conjunction with graphite dust grains were also used to similarly model the mid-IR spectrum of Tc~1, resulting in a worst fit of the red wing of the broad 12\,{\micron} feature (see the Appendix~\ref{sec:ap_1}).
Finally, we stress that preliminary photoionization models with HAC-like grains deposited at 300\,K also provide acceptable good fits to the 12\,{\micron} plateau feature (and dust continuum) in another fullerene PNe such as IC~418 and SMC~16, when playing with the model parameters (e.g., grain size and shape distribution; G{\'o}mez-Mu{\~n}oz et al., in prep.).

\section{Astrophysical implications} \label{sec:implications}

To the best of our knowledge, this is the first time that HAC grains convincingly reproduce the 12\,{\micron} plateau emission feature in a fullerene PN (Tc~1), posing doubts on the generally accepted SiC identification in the literature as the main carrier of this feature. Remarkably, the HAC grains naturally provide an important fraction of the IR dust continuum emission and are consistent with the other UIR features observed such as  the 7\,{\micron} plateau feature (see below). From our {\sc cloudy} models, we determined that $\sim$1.3--3.0\% (with and without graphite, respectively) of the C in Tc~1 is in the form of HAC-like dust. These values are comparable to the amount of C that is estimated to be in the form of fullerenes, such as C$_{60}$, in Tc~1 \citep[$\sim$1.5\%; see][]{Cami2010}.

The laboratory HAC-like grains (H/C$\sim$0.35) used in our models \citep{Duley2012b} mainly contain aliphatic carbon chains (sp$^{3}$ content), which represent 72\% (i.e., aromatic/aliphatic ratio of $\sim$0.4) of the chemical composition, but also polyyne (-C$\equiv$C-)$_{n}$ and cumulenic (-C=C-)$_{n}$ chains. According to \citet{Duley2012b}, these HAC-like grains can have more aromatic rings than aromatic chains. In fact, they suggest that above 10 $\mu$m, the features can be caused by a variety of vibrational modes associated with ring deformation and with CH$_{n}$ groups (i.e., out-of-plane CH vibrations). In particular, the spectral range of the broad 12\,{\micron} plateau emission may contain the features produced by small aromatic rings (pentagons and hexagons) with side chains such as dimethylnaphthalene or dimethylphenanthrene, but also those produced by solo, duo CH, and CH$_{n}$ groups as components of the hydrocarbon matrix \citep{Duley2012b}. Clearly, the broadening of the 12\,{\micron} plateau emission is the result of this coexistence of small aromatic rings and hydrocarbon chains; that is, the mixed aromatic and aliphatic hydrocarbon composition inside the HAC-like material is the key behind the broadening of the 12\,{\micron} feature. We remark that the only other laboratory optical constants available to us correspond to those measured in HAC films deposited at a lower temperature of 77 K (Prof. W. Duley; priv. comm.) and with slightly higher H concentrations \editone{(H/C$\sim$0.5--0.6), which have fewer aromatic rings in its composition \citep{Duley1998}}.
The 12\,{\micron} plateau feature in Tc~1 cannot be reproduced with the 77 K HAC-like sample (i.e., it is absent; G{\'o}mez-Mu{\~n}oz et al., in prep.), suggesting that the aromatic ring/chain ratio, rather than the aromatic/aliphatic ratio, is a key parameter for the formation of the 12\,{\micron} plateau feature.

The presence of HAC-like grains in conjunction with fullerenes have been previously proposed in the literature as proof of the photochemical processing of HAC-like grains for fullerene formation \citep[see e.g.,][]{GarciaHernandez2010,GarciaHernandez2011a,GarciaHernandez2011b,GarciaHernandez2012, BernardSalas2012}. Naturally, our finding in Tc 1 suggests that the formation of fullerenes in PNe circumstellar environments may be related to the photochemical processing of HAC-like grains. However, the only experimental data on the decomposition of HAC-like grains are those from \citet{Scott1997}, where they performed laser ablation (UV radiation-induced decomposition) of HAC films and found PAHs and fullerenes, among other species, but the chemical reactions are presently unknown.
The lower aromatic content compared to more dehydrogenated hydrocarbons in our HAC sample suggest that the main process to generate the precursors to form C$_{60}$, for example, the dehydrogenation of the HAC, is through radiative processing as aliphatic hydrocarbons have been proven to be efficiently dehydrogenated in the ISM by UV radiation \citep[e.g.,][]{MunozCaro2001}.
We speculate that the presence of some small aromatic rings in our 300 K HAC-like sample might act as templates or catalysts for the formation of more complex C$_{n}$ structures when the material is subjected to photochemical processing. More sophisticated laboratory experiments on the photochemical transformation of HAC-like grains at circumstellar conditions would be needed to unveil the details of fullerene formation via chemical reaction routes.

Another obvious implication of our work involves the carriers of the 7\,{\micron} plateau emission feature seen in the fullerene PN Tc~1. Our findings also suggest that HAC-like grains, with H/C=0.35 and mainly composed by aliphatic carbon chains (72\%; see above) with some small aromatic rings, are not a main contributor to such a 7\,{\micron} emission complex and that other still unidentified species should be the carriers. Fullerene-based species are natural candidate species as potential additional emitters at these wavelengths in fullerene PNe. Very recent quantum-chemistry simulations of metallofullerenes (neutral and charged endohedral and exohedral species) do indeed demonstrate that they strongly emit at 6$-$9\,{\micron} \citep{Barzaga2023,Barzaga2023b,Hou2023} with no significant contribution in the 9$-$13\,{\micron} spectral region. 


\section{Concluding remarks} \label{sec:conclusions}

This pilot work leads us to conclude that HAC-like grains are a convincing alternative explanation (i.e., different from the often assumed SiC) for the broad 12\,{\micron} plateau emission feature seen in the fullerene PN Tc~1. In addition, our work suggests that HAC-like grains may be possible carriers of the 12\,{\micron} plateau feature, being likely related to the fullerene formation process (possibly as fullerene precursors) in the circumstellar environment of PNe.

Interestingly, HAC-like grains can also reproduce the far-ultraviolet (FUV) rise seen in the extinction curve (\citealt{Gavilan2016,Gavilan2017}) and very recently detected also in the International Ultraviolet Explorer (IUE) UV spectrum of Tc~1 \citep{GomezMunoz2024}. In this particular case, \citet{GomezMunoz2024} could reproduce the FUV rise toward Tc~1 very well by using HAC-like nanograins (1-15\,nm). Although other carbonaceous materials such as hydrogenated carbon clusters and large PAHs, possibly among others, may also explain the FUV rise in extinction curves \citep[see e.g.,][]{Dubosq2023,Lin2023}; such an independent finding would be consistent with the results presented here.

Yet, we do not fully understand the exact chemical pathways from large HAC-like dust grains toward the most stable fullerene molecules as well as the origin of the other broad plateau emission features seen in fullerene PNe, such as 6$-$9, 15$-$20, and 25$-$35\,{\micron}. The James Webb Space Telescope offers the opportunity to observe these broad features in fullerene PNe at a much higher sensitivity and spectral resolution than \textit{Spitzer}, permitting the variations in their positions and shapes to be studied precisely (even resolving a possible substructure) and providing new additional constraints for future modeling studies. Undoubtedly, more laboratory efforts are needed in order to definitively reveal the details of fullerene formation in PNe circumstellar environments as well as the carriers of the elusive UIR plateau features. Our work strongly encourages more laboratory experiments to obtain the refractive indices $n$ and $k$ (over a large wavelength range, from the UV to the far-IR) of HAC-like dust grains with several structures and a composition (e.g., H/C ratios) as well as at different physical conditions (e.g., temperatures or aromatic ring/chain ratios). Such laboratory measurements (not presently available) would be used to expand the pilot novel work presented here for Tc 1 to a larger sample of fullerene PNe.


\begin{acknowledgements}
      We kindly acknowledge Prof. Walt Duley for supplying us with the laboratory optical constants of HAC-like particles and Dr. Christophe Morisset for earlier work and discussions on the IR spectral modeling, including HAC-like dust, with the photoionization code {\sc cloudy}. We acknowledge the support from the State Research Agency (AEI) of the Spanish Ministry of Science and Innovation (MCIN) under grant PID2020-115758GB-I00/AEI/10.13039/501100011033. This article is based upon work from European Cooperation in Science and Technology (COST) Action NanoSpace, CA21126, supported by COST. This work is based on observations made with the Spitzer Space Telescope, which was operated by the Jet Propulsion Laboratory, California Institute of Technology under a contract with NASA.
\end{acknowledgements}

%
%

\bibliographystyle{aa} 
\bibliography{main} 

\begin{thebibliography}{49}
\expandafter\ifx\csname natexlab\endcsname\relax\def\natexlab#1{#1}\fi

\bibitem[{{Aleman} {et~al.}(2019){Aleman}, {Leal-Ferreira}, {Cami}, {Akras},
  {Ochsendorf}, {Wesson}, {Morisset}, {Cox}, {Bernard-Salas}, {Paladini},
  {Peeters}, {Stock}, {Monteiro}, \& {Tielens}}]{Aleman2019}
{Aleman}, I., {Leal-Ferreira}, M.~L., {Cami}, J., {et~al.} 2019, \mnras, 490,
  2475

\bibitem[{{Arun} {et~al.}(2023){Arun}, {Mathew}, {Manoj}, {Maheswar},
  {Shridharan}, {Kartha}, \& {Narang}}]{Arun2023}
{Arun}, R., {Mathew}, B., {Manoj}, P., {et~al.} 2023, \mnras, 523, 1601

\bibitem[{{Barzaga} {et~al.}(2023{\natexlab{a}}){Barzaga},
  {Garc{\'\i}a-Hern{\'a}ndez}, {D{\'\i}az-Tendero}, {Sadjadi}, {Manchado}, \&
  {Alcami}}]{Barzaga2023}
{Barzaga}, R., {Garc{\'\i}a-Hern{\'a}ndez}, D.~A., {D{\'\i}az-Tendero}, S.,
  {et~al.} 2023{\natexlab{a}}, \apj, 942, 5

\bibitem[{{Barzaga} {et~al.}(2023{\natexlab{b}}){Barzaga},
  {Garc{\'\i}a-Hern{\'a}ndez}, {D{\'\i}az-Tendero}, {Sadjadi}, {Manchado},
  {Alcami}, {G{\'o}mez-Mu{\~n}oz}, \& {Huertas-Rold{\'a}n}}]{Barzaga2023b}
{Barzaga}, R., {Garc{\'\i}a-Hern{\'a}ndez}, D.~A., {D{\'\i}az-Tendero}, S.,
  {et~al.} 2023{\natexlab{b}}, \apjs, 269, 26

\bibitem[{{Bernal} {et~al.}(2019){Bernal}, {Haenecour}, {Howe}, {Zega},
  {Amari}, \& {Ziurys}}]{Bernal2019}
{Bernal}, J.~J., {Haenecour}, P., {Howe}, J., {et~al.} 2019, \apjl, 883, L43

\bibitem[{{Bernard-Salas} {et~al.}(2012){Bernard-Salas}, {Cami}, {Peeters},
  {Jones}, {Micelotta}, \& {Groenewegen}}]{BernardSalas2012}
{Bernard-Salas}, J., {Cami}, J., {Peeters}, E., {et~al.} 2012, \apj, 757, 41

\bibitem[{{Bern{\'e}} \& {Tielens}(2012)}]{Berne2012}
{Bern{\'e}}, O. \& {Tielens}, A. G.~G.~M. 2012, Proceedings of the National
  Academy of Science, 109, 401

\bibitem[{{Blanco} {et~al.}(1988){Blanco}, {Bussoletti}, \&
  {Colangeli}}]{Blanco1988}
{Blanco}, A., {Bussoletti}, E., \& {Colangeli}, L. 1988, \apj, 334, 875

\bibitem[{{Buss} {et~al.}(1990){Buss}, {Cohen}, {Tielens}, {Werner}, {Bregman},
  {Witteborn}, {Rank}, \& {Sandford}}]{Buss1990}
{Buss}, R.~H., J., {Cohen}, M., {Tielens}, A.~G.~G.~M., {et~al.} 1990, \apjl,
  365, L23

\bibitem[{{Cami} {et~al.}(2010){Cami}, {Bernard-Salas}, {Peeters}, \&
  {Malek}}]{Cami2010}
{Cami}, J., {Bernard-Salas}, J., {Peeters}, E., \& {Malek}, S.~E. 2010,
  Science, 329, 1180

\bibitem[{{Cohen} {et~al.}(1985){Cohen}, {Tielens}, \&
  {Allamandola}}]{Cohen1985}
{Cohen}, M., {Tielens}, A.~G.~G.~M., \& {Allamandola}, L.~J. 1985, \apjl, 299,
  L93

\bibitem[{{Dubosq} {et~al.}(2023){Dubosq}, {Pla}, {Dartois}, \&
  {Simon}}]{Dubosq2023}
{Dubosq}, C., {Pla}, P., {Dartois}, E., \& {Simon}, A. 2023, \aap, 670, A175

\bibitem[{{Duley}(1984)}]{Duley1984}
{Duley}, W.~W. 1984, \apj, 287, 694

\bibitem[{{Duley} \& {Hu}(2012)}]{Duley2012b}
{Duley}, W.~W. \& {Hu}, A. 2012, \apj, 761, 115

\bibitem[{{Duley} {et~al.}(1998){Duley}, {Scott}, {Seahra}, \&
  {Dadswell}}]{Duley1998}
{Duley}, W.~W., {Scott}, A.~D., {Seahra}, S., \& {Dadswell}, G. 1998, \apjl,
  503, L183

\bibitem[{{Ferland} {et~al.}(2017){Ferland}, {Chatzikos}, {Guzm{\'a}n},
  {Lykins}, {van Hoof}, {Williams}, {Abel}, {Badnell}, {Keenan}, {Porter}, \&
  {Stancil}}]{Ferland2017}
{Ferland}, G.~J., {Chatzikos}, M., {Guzm{\'a}n}, F., {et~al.} 2017, \rmxaa, 53,
  385

\bibitem[{{Ferrarotti} \& {Gail}(2006)}]{Ferrarotti2006}
{Ferrarotti}, A.~S. \& {Gail}, H.~P. 2006, \aap, 447, 553

\bibitem[{{Garc{\'\i}a-Hern{\'a}ndez}(2012)}]{GarciaHernandez2012b}
{Garc{\'\i}a-Hern{\'a}ndez}, D.~A. 2012, in Planetary Nebulae: An Eye to the
  Future, Vol. 283, 148--155

\bibitem[{{Garc{\'\i}a-Hern{\'a}ndez}
  {et~al.}(2011{\natexlab{a}}){Garc{\'\i}a-Hern{\'a}ndez}, {Iglesias-Groth},
  {Acosta-Pulido}, {Manchado}, {Garc{\'\i}a-Lario}, {Stanghellini}, {Villaver},
  {Shaw}, \& {Cataldo}}]{GarciaHernandez2011b}
{Garc{\'\i}a-Hern{\'a}ndez}, D.~A., {Iglesias-Groth}, S., {Acosta-Pulido},
  J.~A., {et~al.} 2011{\natexlab{a}}, \apjl, 737, L30

\bibitem[{{Garc{\'\i}a-Hern{\'a}ndez}
  {et~al.}(2011{\natexlab{b}}){Garc{\'\i}a-Hern{\'a}ndez}, {Kameswara Rao}, \&
  {Lambert}}]{GarciaHernandez2011a}
{Garc{\'\i}a-Hern{\'a}ndez}, D.~A., {Kameswara Rao}, N., \& {Lambert}, D.~L.
  2011{\natexlab{b}}, \apj, 729, 126

\bibitem[{{Garc{\'\i}a-Hern{\'a}ndez}
  {et~al.}(2010){Garc{\'\i}a-Hern{\'a}ndez}, {Manchado}, {Garc{\'\i}a-Lario},
  {Stanghellini}, {Villaver}, {Shaw}, {Szczerba}, \&
  {Perea-Calder{\'o}n}}]{GarciaHernandez2010}
{Garc{\'\i}a-Hern{\'a}ndez}, D.~A., {Manchado}, A., {Garc{\'\i}a-Lario}, P.,
  {et~al.} 2010, \apjl, 724, L39

\bibitem[{{Garc{\'\i}a-Hern{\'a}ndez}
  {et~al.}(2012){Garc{\'\i}a-Hern{\'a}ndez}, {Villaver}, {Garc{\'\i}a-Lario},
  {Acosta-Pulido}, {Manchado}, {Stanghellini}, {Shaw}, \&
  {Cataldo}}]{GarciaHernandez2012}
{Garc{\'\i}a-Hern{\'a}ndez}, D.~A., {Villaver}, E., {Garc{\'\i}a-Lario}, P.,
  {et~al.} 2012, \apj, 760, 107

\bibitem[{{Gavilan} {et~al.}(2016){Gavilan}, {Alata}, {Le}, {Pino}, {Giuliani},
  \& {Dartois}}]{Gavilan2016}
{Gavilan}, L., {Alata}, I., {Le}, K.~C., {et~al.} 2016, \aap, 586, A106

\bibitem[{{Gavilan} {et~al.}(2017){Gavilan}, {Le}, {Pino}, {Alata}, {Giuliani},
  \& {Dartois}}]{Gavilan2017}
{Gavilan}, L., {Le}, K.~C., {Pino}, T., {et~al.} 2017, \aap, 607, A73

\bibitem[{{Gielen} {et~al.}(2011){Gielen}, {Cami}, {Bouwman}, {Peeters}, \&
  {Min}}]{Gielen2011}
{Gielen}, C., {Cami}, J., {Bouwman}, J., {Peeters}, E., \& {Min}, M. 2011,
  \aap, 536, A54

\bibitem[{{G{\'o}mez-Llanos} {et~al.}(2018){G{\'o}mez-Llanos}, {Morisset},
  {Szczerba}, {Garc{\'\i}a-Hern{\'a}ndez}, \&
  {Garc{\'\i}a-Lario}}]{GomezLlanos2018}
{G{\'o}mez-Llanos}, V., {Morisset}, C., {Szczerba}, R.,
  {Garc{\'\i}a-Hern{\'a}ndez}, D.~A., \& {Garc{\'\i}a-Lario}, P. 2018, \aap,
  617, A85

\bibitem[{{G{\'o}mez-Mu{\~n}oz} {et~al.}(2024){G{\'o}mez-Mu{\~n}oz},
  {Garc{\'\i}a-Hern{\'a}ndez}, {Manchado}, {Barzaga}, \&
  {Huertas-Rold{\'a}n}}]{GomezMunoz2024}
{G{\'o}mez-Mu{\~n}oz}, M.~A., {Garc{\'\i}a-Hern{\'a}ndez}, D.~A., {Manchado},
  A., {Barzaga}, R., \& {Huertas-Rold{\'a}n}, T. 2024, \mnras, 528, 2871

\bibitem[{{Herwig}(2005)}]{herwig2005}
{Herwig}, F. 2005, \araa, 43, 435

\bibitem[{{Higdon} {et~al.}(2004){Higdon}, {Devost}, {Higdon}, {Brandl},
  {Houck}, {Hall}, {Barry}, {Charmandaris}, {Smith}, {Sloan}, \&
  {Green}}]{Higdon2004}
{Higdon}, S.~J.~U., {Devost}, D., {Higdon}, J.~L., {et~al.} 2004, \pasp, 116,
  975

\bibitem[{{Hou} {et~al.}(2023){Hou}, {Lushchikova}, {Bakker}, {Lievens},
  {Decin}, \& {Janssens}}]{Hou2023}
{Hou}, G.-L., {Lushchikova}, O.~V., {Bakker}, J.~M., {et~al.} 2023, \apj, 952,
  13

\bibitem[{{Houck} {et~al.}(2004){Houck}, {Roellig}, {van Cleve}, {Forrest},
  {Herter}, {Lawrence}, {Matthews}, {Reitsema}, {Soifer}, {Watson}, {Weedman},
  {Huisjen}, {Troeltzsch}, {Barry}, {Bernard-Salas}, {Blacken}, {Brandl},
  {Charmandaris}, {Devost}, {Gull}, {Hall}, {Henderson}, {Higdon}, {Pirger},
  {Schoenwald}, {Sloan}, {Uchida}, {Appleton}, {Armus}, {Burgdorf},
  {Fajardo-Acosta}, {Grillmair}, {Ingalls}, {Morris}, \& {Teplitz}}]{Houck2004}
{Houck}, J.~R., {Roellig}, T.~L., {van Cleve}, J., {et~al.} 2004, \apjs, 154,
  18

\bibitem[{{Jones}(2012)}]{Jones2012}
{Jones}, A.~P. 2012, \aap, 542, A98

\bibitem[{{Jones} {et~al.}(2023){Jones}, {{\'A}lvarez-M{\'a}rquez}, {Sloan},
  {Kavanagh}, {Argyriou}, {Law}, {Labiano}, {Patapis}, {Mueller}, {Larson},
  {Bright}, {Klaassen}, {Fox}, {Gasman}, {Geers}, {Glauser}, {Guillard},
  {Nayak}, {Noriega-Crespo}, {Ressler}, {Sargent}, {Temim}, {Vandenbussche}, \&
  {Garc{\'\i}a Mar{\'\i}n}}]{Jones2023}
{Jones}, O.~C., {{\'A}lvarez-M{\'a}rquez}, J., {Sloan}, G.~C., {et~al.} 2023,
  \mnras, 523, 2519

\bibitem[{{Kroto} {et~al.}(1985){Kroto}, {Heath}, {Obrien}, {Curl}, \&
  {Smalley}}]{Kroto1985}
{Kroto}, H.~W., {Heath}, J.~R., {Obrien}, S.~C., {Curl}, R.~F., \& {Smalley},
  R.~E. 1985, \nat, 318, 162

\bibitem[{{Kwok}(2016)}]{kwok2016}
{Kwok}, S. 2016, \aapr, 24, 8

\bibitem[{{Kwok} {et~al.}(2001){Kwok}, {Volk}, \& {Bernath}}]{Kwok2001}
{Kwok}, S., {Volk}, K., \& {Bernath}, P. 2001, \apjl, 554, L87

\bibitem[{{Kwok} \& {Zhang}(2011)}]{Kwok2011}
{Kwok}, S. \& {Zhang}, Y. 2011, \nat, 479, 80

\bibitem[{{Lin} {et~al.}(2023){Lin}, {Yang}, \& {Li}}]{Lin2023}
{Lin}, Q., {Yang}, X.~J., \& {Li}, A. 2023, \mnras, 525, 2380

\bibitem[{{Mu{\~n}oz Caro} {et~al.}(2001){Mu{\~n}oz Caro}, {Ruiterkamp},
  {Schutte}, {Greenberg}, \& {Mennella}}]{MunozCaro2001}
{Mu{\~n}oz Caro}, G.~M., {Ruiterkamp}, R., {Schutte}, W.~A., {Greenberg},
  J.~M., \& {Mennella}, V. 2001, \aap, 367, 347

\bibitem[{{Murga} {et~al.}(2022){Murga}, {Akimkin}, \& {Wiebe}}]{Murga2022}
{Murga}, M.~S., {Akimkin}, V.~V., \& {Wiebe}, D.~S. 2022, \mnras, 517, 3732

\bibitem[{{Otsuka} {et~al.}(2014){Otsuka}, {Kemper}, {Cami}, {Peeters}, \&
  {Bernard-Salas}}]{Otsuka2014}
{Otsuka}, M., {Kemper}, F., {Cami}, J., {Peeters}, E., \& {Bernard-Salas}, J.
  2014, \mnras, 437, 2577

\bibitem[{{Perea-Calder{\'o}n} {et~al.}(2009){Perea-Calder{\'o}n},
  {Garc{\'\i}a-Hern{\'a}ndez}, {Garc{\'\i}a-Lario}, {Szczerba}, \&
  {Bobrowsky}}]{PereaCalderon2009}
{Perea-Calder{\'o}n}, J.~V., {Garc{\'\i}a-Hern{\'a}ndez}, D.~A.,
  {Garc{\'\i}a-Lario}, P., {Szczerba}, R., \& {Bobrowsky}, M. 2009, \aap, 495,
  L5

\bibitem[{{Rouleau} \& {Martin}(1991)}]{Rouleau1991}
{Rouleau}, F. \& {Martin}, P.~G. 1991, \jrasc, 85, 201

\bibitem[{{Scott} {et~al.}(1997){Scott}, {Duley}, \& {Pinho}}]{Scott1997}
{Scott}, A., {Duley}, W.~W., \& {Pinho}, G.~P. 1997, \apjl, 489, L193

\bibitem[{{Sellgren} {et~al.}(2010){Sellgren}, {Werner}, {Ingalls}, {Smith},
  {Carleton}, \& {Joblin}}]{Sellgren2010}
{Sellgren}, K., {Werner}, M.~W., {Ingalls}, J.~G., {et~al.} 2010, \apjl, 722,
  L54

\bibitem[{{Speck} {et~al.}(2009){Speck}, {Corman}, {Wakeman}, {Wheeler}, \&
  {Thompson}}]{Speck2009}
{Speck}, A.~K., {Corman}, A.~B., {Wakeman}, K., {Wheeler}, C.~H., \&
  {Thompson}, G. 2009, \apj, 691, 1202

\bibitem[{{van Hoof} {et~al.}(2004){van Hoof}, {Weingartner}, {Martin}, {Volk},
  \& {Ferland}}]{vanHoof2004}
{van Hoof}, P.~A.~M., {Weingartner}, J.~C., {Martin}, P.~G., {Volk}, K., \&
  {Ferland}, G.~J. 2004, \mnras, 350, 1330

\bibitem[{{Werner} {et~al.}(2004){Werner}, {Roellig}, {Low}, {Rieke}, {Rieke},
  {Hoffmann}, {Young}, {Houck}, {Brandl}, {Fazio}, {Hora}, {Gehrz}, {Helou},
  {Soifer}, {Stauffer}, {Keene}, {Eisenhardt}, {Gallagher}, {Gautier}, {Irace},
  {Lawrence}, {Simmons}, {Van Cleve}, {Jura}, {Wright}, \&
  {Cruikshank}}]{Werner2004}
{Werner}, M.~W., {Roellig}, T.~L., {Low}, F.~J., {et~al.} 2004, \apjs, 154, 1

\bibitem[{{Zhang} \& {Kwok}(2011)}]{Zhang2011}
{Zhang}, Y. \& {Kwok}, S. 2011, \apj, 730, 126

\end{thebibliography}

\begin{appendix}

\section{Modeling the Tc~1 mid-IR spectrum using $\alpha$-SiC grains}
\label{sec:ap_1}

\begin{figure}
    \centering
\includegraphics[width=0.99\columnwidth]{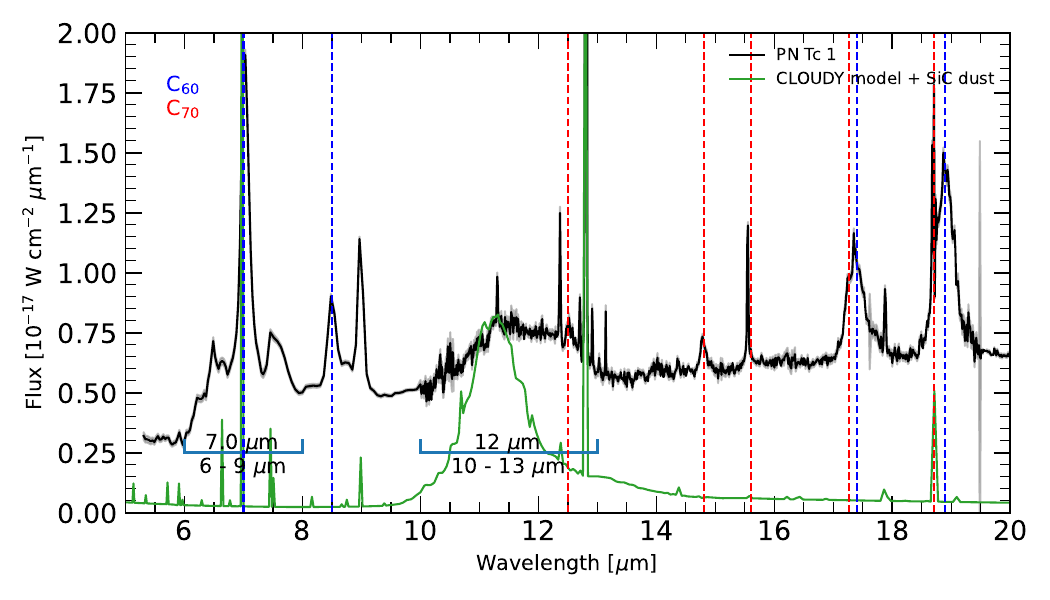}
\includegraphics[width=0.99\columnwidth]{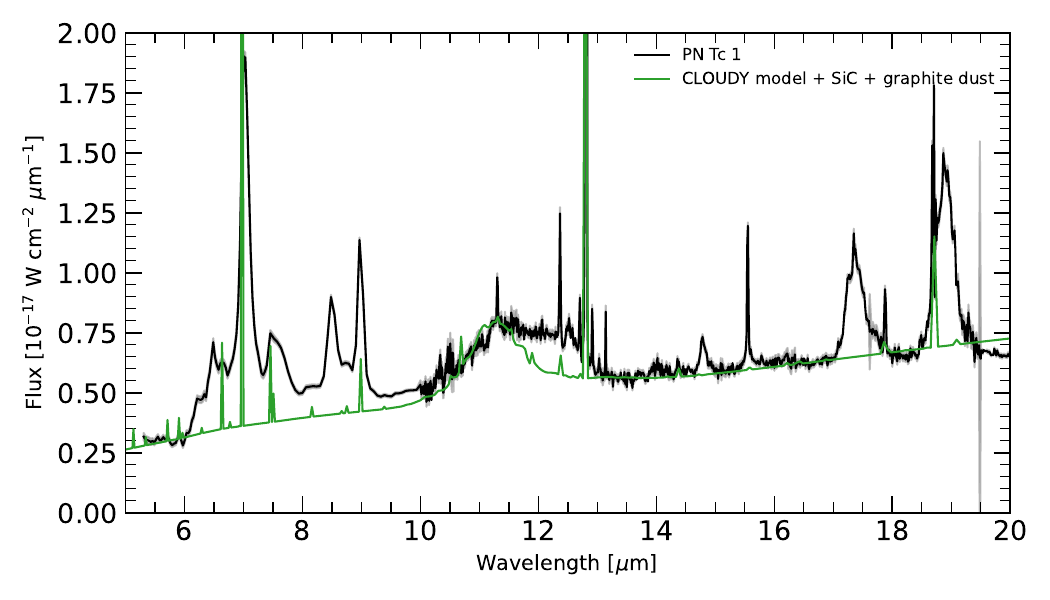}
\caption{Spitzer mid-IR spectrum of the PN Tc~1 (black line) compared with the best photoionization model spectrum including $\alpha$-SiC dust grains only (green line ) are shown (top panel). The C$_{60}$ (blue line) and C$_{70}$ (red line) emission bands and UIR plateau emission features are indicated. We note that the narrow emission features seen in the best model spectrum (green line) are just atomic nebular emission lines, whose detailed modeling is out of the scope of the present work. The best photoionization model including $\alpha$-SiC grains in conjunction with graphite dust grains (green line) is also shown for comparison (bottom panel).
\label{fig:ap_only_sic}}
\end{figure}

For comparison, Fig.~\ref{fig:ap_only_sic} shows the mid-IR spectrum of Tc~1 together with the best {\sc cloudy} model spectrum obtained using $\alpha$-SiC grains \citep[as in][for the case of the PN IC~418]{GomezLlanos2018} with the same shape (spherical) and grain size distribution as when using HAC-like grains. As can be seen from the figure, when using only $\alpha$-SiC (top panel), the best model spectrum only fits the weak 11.3\,{\micron} feature seen in Tc~1 and not the entire broad 12\,{\micron} feature. On the other hand, when using $\alpha$-SiC+graphite (bottom panel), the dust continuum is very well fitted but the red wing of the broad 12\,{\micron} emission is poorly reproduced. 

We remark that it is already well known that much better fits with $\alpha$-SiC can be obtained when playing with very different extreme shapes and grain size distributions \citep[e.g., the case of PN IC 418;][]{GomezLlanos2018} but here we are exploring if there is any other contribution to or explanation for the broad 12\,{\micron} plateau feature in PNe with fullerene-dominated IR spectra such as Tc~1, that is, different from the often assumed SiC.

It is important to mention that the 12\,{\micron} feature in IC~418 has different characteristics (position, width, and emission strength) than in Tc 1 \citep{Otsuka2014}. \citet{GomezLlanos2018} needed to use very different shapes (spheres and five extreme ellipsoid types) and grain size distributions along the nebula in order to fit the red wing of the 12\,{\micron} feature with $\alpha$-SiC. Also, the IR spectrum of IC 418 is contaminated by other species (e.g., PAHs), displaying weaker C$_{60}$ bands and a different fullerene and dust temperature than Tc~1.

\end{appendix}

\end{document}